\newcommand {\sci}[2]    {\mbox{\ensuremath{ #1 \! \cdot \! 10^{#2} }}}
\newcommand {\scix}[1]   {\mbox{\ensuremath{ 10^{#1} }}}
\newcommand {\defuni}[1] {\ifmmode \mathrm{#1} \else $\mathrm{#1}$ \fi}
\newcommand {\um}[1]     {\defuni{\; #1}}

\newcommand {\degr}      {\ensuremath{^{\circ}}}
\newcommand {\SEUSO}     {${\mathcal S}\mathrm{-\textit{EUSO}}$\xspace}
\newcommand {\lumin}     {\um{photons \cdot m^{-2} \cdot s^{-1} \cdot sr^{-1} }}

\newcommand{\ml}{\begin{list}{$\bullet$}{\itemsep=0pt\parsep=0pt\topsep=0pt\leftmargin=0pt}}
\newcommand{\el}{\end{list}}
\documentclass[12pt]{iopart}
\usepackage{iopams}  
\usepackage{graphicx}
\usepackage{xspace}
\usepackage{url}
\begin{document}

\title[Ultra high energy observation from space: \SEUSO]
{Observing Ultra High Energy Cosmic Particles from Space: \SEUSO , the Super-Extreme Universe Space Observatory Mission}

\author{A. Santangelo$^1$ and A. Petrolini$^2$}

\address{
1) Institute f\"ur Astronomie und Astrophysik, Kepler Center for Astro and Particle Physics, Eberhard Karls Universit\"at T\"ubingen, Sand 1, DE-72076, T\"ubingen, Germany.

2) Dipartimento di Fisica dell'Universit\`a di Genova and Istituto Nazionale di
Fisica Nucleare (INFN), Sezione di Genova, Via Dodecaneso 33, I-16146, Genova, Italia.}

\ead{andrea.santangelo@uni-tuebingen.de}

\begin{abstract}
The experimental search for ultra high energy cosmic messengers,
from $E\sim 10^{19}$~eV to beyond $E\sim 10^{20}$~eV, at the very end of
the known energy spectrum, constitutes an extraordinary opportunity to
explore a largely unknown aspect of our universe. Key scientific goals
are the identification of the sources of ultra high energy particles,
the measurement of their spectra and the study of galactic and local
intergalactic magnetic fields. Ultra high energy particles might, also,
carry evidence of unknown physics or of exotic particles relics of the
early universe.

To meet this challenge a significant increase in the integrated exposure is required. This implies a new class of experiments with larger acceptances and good understanding of the systematic uncertainties. Space based observatories can reach the instantaneous aperture and the integrated exposure necessary to systematically explore the ultra high energy universe.

In this paper we focus on the \textit{Super Extreme Universe Space
Observatory}--\SEUSO, a mission concept developed in the framework of
the first Announcement of Opportunity of the "Cosmic Vision 2015-2025"
program, the long term science plan of the European Space Agency. \SEUSO
will observe from space, in a free flyer configuration, the extensive
air showers produced by ultra high energy primaries which traverse the
Earth atmosphere.  From a variable altitude orbit of $800 \div1100
\um{km}$, \SEUSO will have an instantaneous geometrical aperture of $ {A}_{geo} \geq
2\times10^6$~km$^2$~sr with an estimated duty cycle in the range
$10\div20$\%.

In this paper, after briefly summarising the science case of the
mission,  we describe the scientific goals and requirements of the
\SEUSO concept. We then introduce the \SEUSO observational approach and
describe the main instrument and mission features. We conclude
discussing the expected performance of the mission.

\end{abstract}

\pacs{90.00, 98.70Sa, 95.55Vj,-----------------------------------------------}
\vspace{2pc}
\noindent{\it Keywords---------------------------------}: Article preparation, IOP journals
\submitto{\NJP-------------------------------}

\maketitle

\tableofcontents

%\clearpage

%================================================================================
\section{Introduction}
\label{pa:intro}

Observations of cosmic particles at ultra high energies, from a few
$10^{19}$~eV to beyond $10^{20}$ eV, are an
extraordinary opportunity to explore this yet largely unknown
universe and present us a tremendous experimental challenge. It is expected that observations of cosmic rays and neutrinos at ultra high energies will provide entire new information on the sources and on the
physical mechanisms capable to accelerate these extreme messengers to
macroscopic energies. Moreover, these messengers might also carry evidence of unknown physics
or of exotic particles, relics of the early Universe. To carry out such an ambitious program high statistics and high quality observations are needed. The very low flux of these particles, about one particle per km$^{-2}$sr$^{-1}$~millennium$^{-1}$ at energies $E\geq10^{20}$~eV~\cite{Auger1},
requires experiments with large acceptances and good understanding of systematic uncertainties. 
The \textit{Super-Extreme Universe Space Observatory},
\SEUSO~\cite{bi:SEUSO}, is a space based mission to explore the universe
through the study of ultra high energy cosmic particles.

\SEUSO will observe from space, in a free flyer configuration, the
extensive air showers (EAS) produced by ultra high energy cosmic rays
which traverse the Earth atmosphere. Using a target
volume and instantaneous geometrical aperture far greater than what achievable from
ground, \SEUSO is expected to obtain accurate measurements of the
nature, energy and arrival direction of the primary particles.

The ground-based Pierre Auger Observatory~\cite{Watson08} and the
JEM-EUSO~\cite{bi:JEMEUSO} space mission will hopefully provide in the
near future solid bases for the beginning of particle astronomy. However
only a large innovative space-based next-generation mission, which
aims at an instantaneous geometrical aperture of the order of $
\mathcal{A} \approx 10^6 $~km$^2$~sr, can increase by a few orders of
magnitude, the statistics of events with $E\geq10^{20}$~eV, allowing
the identification of the sources of ultra high energy particles.

In this paper we first describe (section~\ref{pa:science}) the scientific reasons at the
base of the \SEUSO concept. We then introduce the \SEUSO observational
approach (section~\ref{pa:approach}) and the instrument and mission
features (section~\ref{pa:apparatus}). The requirements and
expected performance are discussed in section~\ref{pa:perfo}.

The \SEUSO concept has been developed in the framework of the first Announcement Opportunity of the European Space Agency "Cosmic Vision 2015-2025" program, the long term science plan of the Agency.
More than one hundred scientists from forty research groups from Europe, Russia, US and Japan participated to the proposal. 

%================================================================================
\section{The \SEUSO science case}
\label{pa:science}

 Experimenters routinely observe atmospheric showers from particles whose energies reach macroscopic values up to about a few tens of Joules. This dwarfs energies achieved in particle accelerators by about eight orders of magnitude in the detector frame (fixed target experiments) and three orders of magnitude in the centre of mass (collider experiments). Explanations range from conventional shock acceleration in extreme environments to particle physics beyond the Standard Model and processes taking place at the earliest moments of our universe~\cite{Blasireview}. 

Ultra high energy cosmic particles are thought to be coming from extra-galactic
distances. Propagation in largely unknown galactic and extra-galactic magnetic
fields deflects trajectories of charged cosmic rays, limiting proton astronomy
to $E>10^{19}$~eV. On the other hand, the Greisen Zatsepin and Kuz'min effect
(GZK)~\cite{GZK} makes the Universe opaque to proton energies of
$E>5\times10^{19}$~eV. Shortly after the discovery of the CMB, Greisen and
Zatsepin \& Kuz'min independently predicted that pion-producing interactions of cosmic ray protons with CMB photons of target density $\sim400$~cm$^{-3}$
would produce a cut-off in their spectrum at energies greater
than$E\sim5\times10^{19}$~eV, when the pion production threshold is reached. The
reaction $p\gamma \rightarrow\Delta^{+}\rightarrow p \pi^0 // n \pi^{+}$ will
quickly slow down the proton and lead to an effective attenuation length of
about 50 Mpc for a proton of $10^{20}$~eV.  Due to the GZK effect a "flux suppression"
is expected in the spectrum~\cite{Stecker68} which makes their detection
difficult. 

%================================================================================
\subsection{The current observational scenario}
Still ultra high energy cosmic rays exist. After the pioneering detection, back in the '60s, of the first event with the Volcano Ranch Array by J. Linsley~\cite{Linsley63}, ultra high energy particles have been detected by several independent ground-based experiments, including Haverah Park, Yakutsk, AGASA, Fly's Eye, HiRes and recently AUGER (for an historical review see~\cite{NaganoWatson2000}). Up to date a maximum energy of $\sim3.2\times10^{20}$~eV has been reported in literature~\cite{Bird93}. 

The observation scenario has been the subject of an intense debate in the last years: the flux and spectral shape measured by the AGASA observatory~\cite{Takeda03} did not show evidence for a GZK feature, and did not agree with the one observed by the HiRes experiment~\cite{Abu04}. This puzzling scenario was clarified by the measurement of the Pierre AUGER Observatory which together with HiRes reported definitive evidence of the GZK effect~\cite{HiRes08, Auger1}. 

A second point of discrepancy was the small scale clustering of events. Small-scale anisotropies (six pairs and 1 triplet for events with $E\geq5\times10^{19}$~eV and within the $2.5^{\circ}$ AGASA angular resolution) were observed by AGASA and interpreted as evidence for compact sources of ultra high energy cosmic rays~\cite{Takeda99}. These findings were not confirmed by HiRes~\cite{Abbasi04} even if a cluster of five events from the combined published AGASA-HiRes data set was reported by Farrar et al. (2005)~\cite{Farrar05}.  The breakthrough in the field came again with AUGER's discovery of a statistical correlation between the highest energy 27 events ($E\geq5.7\times10^{19}$~eV) and the anisotropically distributed galaxies in the 12th Veron-Cetty \& Veron catalogue of active galactic nuclei (AGN)~\cite{Auger2}. 
%================================================================================
\subsection{Science goals}
The seminal measurement of AUGER signs the beginning of charged particle astronomy. However it does not prove that AGNs are the sources of ultra high energy cosmic rays. Any class of sources which correlate with large scale distribution of matter might be a possible candidate population. AUGER's events show correlation with IRAS PSCz sources~\cite{Kashi08} and with HI emitting galaxies~\cite{Ghisellini08}. To understand which are the sources of these events and to discriminate among the various competing models of acceleration the identification of the sources and the measurement of the spectra of the single sources are crucial. This research program requires exposures a few orders of magnitude larger than the southern site of the AUGER observatory. 

Cosmic rays are mainly charged particles, and therefore they can be used to map galactic and local intergalactic magnetic fields. Protons with $E\geq6\times10^{19}$~eV are deflected by $\sim1^\circ$ traversing $\mu$G (nG) field over a kpc (Mpc). To map the local magnetic field is necessary to measure deflections in a 4$\pi$ coverage of the sky. Full sky coverage of ultra high energy particle flux at high statistics can identify the sources and measure deflections, therefore mapping the local magnetic environment. This has strong implications in cosmology and astrophysics. 

Although charged particle astronomy is at the core of the science case of future cosmic ray observatories other observational windows can be opened by such enterprises. The neutrino Universe is at high and ultra high energies still unexplored. Neutrinos have the advantage over charged cosmic rays of being electrically neutral and not deflected by magnetic fields. Ultra high energy neutrinos point back to the point of their creation. Due to their low interaction cross section, detection of astrophysical neutrinos demands an extraordinarily large volume. \SEUSO will significantly increase the target volume compared to current or planned generation experiments, enabling the exploration of the neutrino universe~\cite{Berezinsky05}. 
Moreover, measurements of neutrino-nucleon cross sections by comparing the rate of horizontal and up-going air showers induced by neutrinos have been discussed in literature~\cite{Kusenko2000, Fargion1999}. In this respect Palomares-Ruiz et al., (2006) have conducted a detailed analysis of the acceptances for space and ground based detectors, finding that the rate of Earth-skimming neutrino induced showers is much higher when observed over the ocean from space than observed from the ground~\cite{Palomares2006}.

Eventually we mention that, as demonstrated by AUGER~\cite{Augerphoton}, another
window that large aperture observatories can open is the direct detection of
photons above the CMB attenuation. 

%================================================================================
\subsection{Planning observatories for the future: why from space?}

AUGER studies will be extended to the northern hemisphere with a second site
consisting of 4.000 detector stations, to be deployed in Colorado, US. AUGER
North aims at reaching a geometrical area of
$A_{geo}\sim 2\times10^4$~km$^2$. This converts into an effective aperture of
more than $A_{eff}\sim45.000$~km$^2$~sr.

In any post-AUGER scenario observations from space are likely to be
essential. Space-based observatories can in fact reach a practical instantaneous geometrical
aperture up to $A_{eff}\sim2.5\times10^6$~km$^2$~sr that translates in a target
mass of more than $10^{12}$~ton, with full sky coverage. Assuming a duty cycle
$\eta\sim10 \div 20\%$ and an operation time of about five years this converts
into an exposure of $A_{exp}\sim(1.2 \div 2.0)\times10^6$~km$^2$~sr~yr. 

The original idea to observe, by means of space-based devices looking at Nadir
nighttime, the fluorescence light (300$\div$400~nm) produced by an EAS proceeding in the atmosphere, goes back to 1979, when J. Linsley firstly suggested the "SOCRAS" concept~\cite{Linsley82}. The "SOCRAS" concept triggered the AIRWATCH program in Europe, which after a few years led to EUSO. 

The \textit{Extreme Universe Space Observatory (EUSO)}, was originally an ESA
lead international mission designed for the Columbus module on the International
Space Station (ISS, at 430 km), characterised by an $A_{exp}\sim(1.3 \div
3.2)\times10^5$~km$^2$~sr~yr. The phase A study was successfully completed in
2004. Although EUSO was found technically ready, ESA did not continue the
mission mainly due to programmatic uncertainties related to the ISS. The EUSO
mission concept has been recently re-oriented to JEM-EUSO. The JEM-EUSO space
observatory, led by Japan, is an EUSO-like concept to be accommodated on the
Japanese Exposure Module (JEM) of ISS. The mission is currently in its phase
A/B. Several aspects like the optics, the sensors quantum efficiency and the
trigger scheme have been improved with respect to EUSO. The instantaneous geometrical aperture
of the mission is $A_{geo}\sim1.5\times10^5$~km$^2$ which converts to
$A_{exp}\sim(2.1 \div 4.3)\times10^5$~km$^2$~sr~yr (5 years in
operation)~\cite{Takahashi09}. 

In 2007, following a call for opportunities of
ESA in the framework of the scientific program of the agency for next decade
(the "Cosmic Vision Program 2015-2025"), a proposal for a large aperture
\textit{free-flyer observatory} for ultra high energy studies was submitted to the Agency: the
\SEUSO mission~\cite{ bi:SEUSO, Santangelo2005}. Thanks to its planned
higher orbit, \SEUSO will have an instantaneous geometrical aperture larger by a factor of six with respect to JEM-EUSO. Moreover, because of its larger optics and better photon detection efficiency, \SEUSO is expected to reach a significant lower thresholds than JEM-EUSO. The higher quality signal could increase the duty cycle. We therefore expect that \SEUSO will collect a factor of ten more events than JEM-EUSO.

%================================================================================
\subsection{Scientific Objectives}
\SEUSO is expected to detect several thousands of events at $E \geq 5\times10^{19}\um{eV}$.
The main science objectives of \SEUSO are:
\begin{enumerate}
\item
The extension of the measurement of cosmic ray spectrum beyond $ E \approx 10^{20} \um{eV}$, reaching $E \approx 10^{21}\um{eV}$.
\item
The detailed map of the arrival distribution of cosmic rays extended to the entire sky. 
The localisation and identification of "compact" sources. The map of the deflections. 
\item
The study of the spectra of individual sources.
\item
Composition studies in the energy range $E\sim10^{19}\div 10^{20}$.
\end{enumerate}
Other scientific objectives of the mission are: 
\begin{enumerate}
\item The measurement of the flux of "compact" and diffuse sources 
of ultra high energy neutrinos.
\item The search and identification of horizontal and skimming showers induced by $\tau$ neutrinos.  
\item
The measurement of the flux of the ultra high energy photon component.
\end{enumerate}
Other scientific objectives specific of atmosphere science are not discussed here. We refer the reader to the \SEUSO proposal~\cite{bi:SEUSO} for details.
 %================================================================================
\subsection{Scientific Requirements}
The scientific objectives summarised above dictate the following scientific
requirements:

\begin{enumerate}
\item
Low energy threshold (flat efficiency plateau $\sim 100$\%) 
at $E_{th} \simeq 10^{19} \um{eV} $.
\item
Statistical uncertainty on the energy measurement: 
$ \Delta E/E \approx 0.1 \; @ \; 10^{19} \um{eV} $.
\item
Instantaneous geometrical aperture averaged on orbit:
$ {A}_{geo} \geq 2\times10^6$~km$^2$~sr. The duty cycle is not included:
current estimates, based on the EUSO studies, indicate $ \eta \approx (0.1 \div 0.2) $. 
\item
An angular granularity corresponding to $ \Delta \ell \approx 1 \um{km} $ at the
Earth.
\item
An average angular resolution on the reconstructed direction of
$ \Delta\chi \approx 1\degr \div 2\degr \; @ \; 10^{20} \um{eV} $;
the angular resolution strongly depends on the EAS zenith angle: 
inclined EAS will have a better than average angular resolution.
\item
An average angular resolution on the reconstructed particle direction of
$ \Delta X_{max} \approx 20\div50$~gr~cm$^{-2}$ @ $10^{20} \um{eV} $;
resolution on $X_{max}$ depends also on the EAS zenith angle.
\end{enumerate}

%================================================================================
\section{The observational approach}
\label{pa:approach}

\begin{figure}[ht]
\begin{center}
\includegraphics[width=5.6truecm]{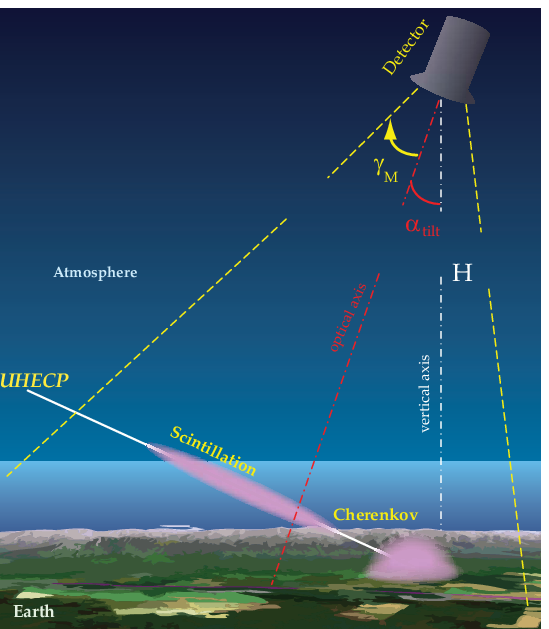}
\includegraphics[width=8.3truecm]{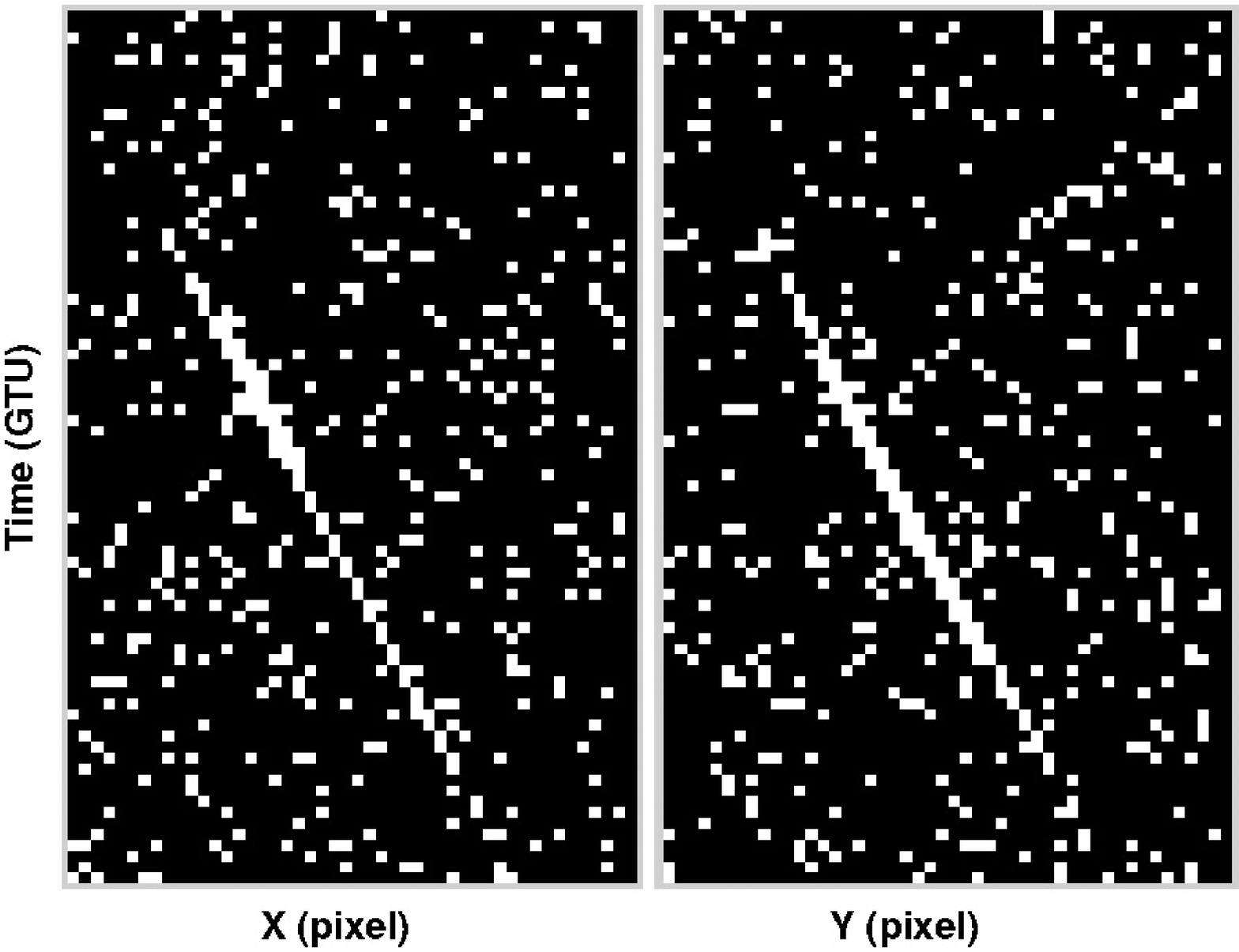}
\end{center}
\caption{\small Left Panel) The \SEUSO observational approach (original figure from~\cite{bi:ArXiv}). Right panel) The formation of the tracks in the X, Y vs. time planes (original from~\cite{bi:RedBook}). The typical size of the pixel correspond to $\Delta \ell \approx 0.7 $~km on Earth. The typical value for the Gate Time Unit (GTU) is $2.5 \mu s$ }
\label{fi:approach}
\end{figure}

\SEUSO uses the Earth atmosphere, viewed from space at night, as a calorimeter to
measure the nature, energy and arrival direction of the ultra high energy particle induced
EAS. The \SEUSO observational method is shown in figure~\ref{fi:approach}. It is
based on the measurement of fluorescence photons produced by the EAS as it
progresses through the atmosphere.

%================================================================================
\subsection{The \SEUSO observational technique}
A hadronic particle (interaction length $\sim40$~g~cm$^{-2}$ at
$E\sim10^{20}$~eV) penetrating the Earth atmosphere generates a shower of
secondary particles. The number of these secondary particles, largely dominated
by electrons/positrons, reaches  at Òshower maximumÒ $N\geq10^{11}$,
proportional to the energy of the primary particle. The total energy carried by
the charged secondary particles ($\sim0.5\%$) is converted into fluorescence
photons through the excitation of the air N$_{2}$ molecules. The fluorescence
light is isotropic and proportional, at any point, to the number of charged
particles in the EAS. The total amount of light produced is proportional to the
primary particle energy and the shape of the EAS profile (in particular the
atmospheric depth of the EAS maximum) contains information about the primary
particle identity. The fluorescence yield in air, $Y_{air}$, in the
\mbox{$330\div400$~nm} wavelength range, is about
\mbox{$Y_{air} \approx 4.5$} photons per charged particle
per meter at \mbox{$h \lesssim 20 \um{km}$}, depending, in a known way, from
altitude, pressure, temperature and air composition~\cite{Kakimoto,
  Nagano}. Uncertainties are of $\sim15$\%. 
The main emission lines are located near the three wavelengths
$337\um{nm}$, $357 \um{nm}$ and $391\um{nm}$. The fluorescence emission of the
shower is rather constant for $h<15$~km and appears as a thin luminous disk of
radius of the order of 0.1 km and depth of the order a few meters.  It moves
through the 
atmosphere at the speed of light. 

A highly beamed Cherenkov radiation is generated as well by the ultra-relativistic particles in the EAS and partly scattered by the atmosphere itself.
The additional observation of the diffusely reflected Cherenkov light (reflected
either by land, sea or clouds) provides additional information, such as the
landing point and timing, useful to improve the EAS reconstruction. 
It greatly helps in determining the EAS parameters. While the amount of observed Cherenkov photons depends
on the reflectance and geometry of the impact surface, the
directionality of the Cherenkov beam provides a precise extrapolation
of the EAS to the first reflecting surface. The Cherenkov light will
be seen as a bunch of photons coming from a limited region in a short
time interval. The total number of Cherenkov photons generated in the
\mbox{$330 \um{nm} \div 400 \um{nm}$} wavelength range, is roughly of the same
order of magnitude as the number of generated fluorescence photons. Cherenkov light scattered at high angles during the EAS development can reach \SEUSO by multiple scattering. 

Typically, for a $10^{20}$~eV EAS, a few thousand photons will reach the \SEUSO detector. As the \SEUSO telescope has a mirror system associated to a fast counting, pixelized photo-detector on the focal surface, \SEUSO will detect not only the number of arriving photons but also their direction and time of arrival. It's the observation of this specific space-time correlation that identifies, very precisely, the presence of an ultra high energy shower (see figure 1). 

%===============================================================================

\subsection{Atmosphere, background and the duty cycle}
The atmosphere acts as signal generator
(fluorescence and Cherenkov light), as a signal attenuator
(scattering and absorption) and as source of background.

The atmosphere is relatively transparent down to $\lambda \approx 330
\um{nm}$, where the ozone absorption becomes strong. Preliminary
simulations show that for typical cloud-less atmosphere models the vertical transmission
coefficient from Earth surface to \SEUSO, is $t \gtrsim 0.3$,
in all the interesting wavelength range. Of course multiple scattering
will generate some background. The main atmospheric components
affecting the signal transmission are Rayleigh and Mie scattering,
ozone absorption (severe up to $\lambda\simeq 330\um{nm}$), and the presence of clouds
(affecting either signal transmission and EAS characterisation). Losses are dominated by Rayleigh scattering. Real time measurements of these components are mandatory to control \SEUSO systematics.

The main background component is the random night-glow background from the Earth
albedo.  A second relevant component is due to the light from air-glow, which
has been measured by several experiments~\cite{Barbier, Garipov}.  The random background has also
contributions from zodiacal light, star light and artificial scattered light. 
In addition many different sources can give rise to background events that
must be discriminated against cosmic ray events. They include man-made lights, auroras, natural
photo-chemical effects (in atmosphere, sea and land), low-energy cosmic
radiation. 
The signal associated with these background sources develops typically in a time-scale of the order of ms to be compared with the tens-hundreds of \um{\mu s} time duration of the ultra high energy shower signal. Therefore these spurious events can be discriminated and rejected through studies of the kinematic of the tracks. 

Based on the known available data we estimate for \SEUSO a
conservative value of $ (3 \div 10)\times10^{11}$~photons~m$^{-2}$~s$^{-1}$~sr$^{-1}$ in the
wavelength range \mbox{$330 \um{nm} \div 400 \um{nm}$}~\cite{bi:SEUSO}.

A precise value for the duty cycle can be hardly estimated and 
dedicated measurements from space would be crucial. The duty cycle depends on the amount of background level that can be accepted by \SEUSO without compromising data reconstruction.  This is of course function of the energy. Partial moon-light may, in some instances, not prohibit the
\SEUSO detector from observing very high energy EAS. We preliminarly estimate the duty cycle to be $\eta \simeq (0.1 \div 0.2 )$. More details can be found in~\cite{Berat2003}.  
%===============================================================================

 \subsection{The observation of EAS from space}

\SEUSO will observe the Earth atmosphere during night-time and low moon-light condition,
by looking down to nadir with large aperture and large field of view optics, focusing the image onto a
highly pixelized and fast photo-detector. The spatial and temporal development of the EAS in the atmosphere are therefore recorded. 

The sufficiently fast response of \SEUSO allows to determine
the direction of the cosmic ray primary by means of one single observation
point. An EAS will be seen as a point moving inside the field of view
with a direction and an angular velocity depending on the EAS
direction. The EAS velocity can be decomposed
into its parallel and perpendicular components with respect to the line
of sight joining \SEUSO to any suitable point of the EAS. 
As the speed of the EAS is known (equal to the speed of light) the EAS direction
can be easily determined from kinematics considerations.

Several main features of the space based observational can be anticipated.
\SEUSO will have a large instantaneous geometrical aperture of the order of  $A_{eff}\sim2.5\times10^6$~km$^2$~sr that translates in a target mass of more than $10^{12}$~ton. The geometrical acceptance of any space experiment is well defined by the field of view. However the fact that the observed part of the atmosphere is continuosly changing requires an atmosphere monitoring device.  For space-based experiments the distance of the EAS, which develops in the
lowest part of the atmosphere, is an approximatively known quantity, in contrast to ground based experiments where a strong correction due to the proximity effect is mandatory.  Photon propagation from the EAS to the experimental apparatus occurs, for space experiments, through the less dense part of the atmosphere. Moreover the effect of Mie scattering is considerably reduced as the aerosols are
mainly concentrated in the lowest part of the atmosphere. Contamination of the fluorescence light by direct Cherenkov light is small for space experiments, unlike ground-based detectors.
The EAS development can be registered in position and time
when the EAS hits land or sea by detecting the diffusely reflected Cherenkov landmark. The same applies when reflection occurs by a cloud layer, provided that the knowledge of the height of the cloud layer is known.  
All sky coverage is possible with one single experiment, depending on the orbit.
Observation of deeply penetrating EAS, from primary particles
interacting deeply in the atmosphere (like neutrinos), is possible by the
direct observation of the EAS development and starting point.
%================================================================================
\section{The experimental apparatus}
\label{pa:apparatus}

%================================================================================
\subsection{Architecture of the instrument}

The \SEUSO mission is an enlarged and improved \textit{free-flyer} version of the
former EUSO mission concept. It surpasses EUSO by a much larger aperture and by exploiting novel technologies. Table~\ref{ta:params} summarises the main features of the instrument and of the mission.

\SEUSO consists of the following parts:
\begin{itemize}
\item
Main telescope operating in the near-UV. It's a large aperture, large field of view fast, pixelized instrument working in single photon counting. Its parts are:
\begin{itemize}
\item
A main reflective deployable optics consisting of:
\begin{itemize}
\item
the main mirror: a large, lightweight, segmented, nearly spherical, deployable
mirror;
\item
the corrector plate on the entrance pupil (deployable as well);
\item
the optical filters;
\item
active control mechanisms for both the mirror and the corrector plate;
\item
supporting structure and ancillary parts.
\item
the optical baffle.
\end{itemize}
\item
The photo-detector (PD) on the FS of the optics made of:
\begin{itemize}
\item
the photo-sensors; arrays of GAPD as baseline;
\item
the light-collection system, to increase the acceptance of the photo-sensors;
\item
the front-end electronics;
\item
the back-end, trigger and on-board data-handling electronics.
\end{itemize}
\end{itemize}
\item
The Atmospheric Monitoring System:
\begin{itemize}
\item
a dedicated LIDAR;
\item
an infrared camera;.
\end{itemize}
\end{itemize}
Other crucial parts of the instrument are the monitoring, alignment and calibration system; the central control unit providing the intelligence to all the systems; The control and power systems. These are discussed in details in~\cite{bi:SEUSO}.

\subsubsection{The Optics.}
Mission constrains require the optics to be lightweight and deployable. 
The proposed optics baseline is based on a catadioptric design, in a Schmidt configuration. The baseline approach has been investigated in the context  of an ESA study for an Earth-looking LIDAR telescope~\cite{Optic1,Optic2}. A similar solution was studied by NASA for OWL~\cite{OWL}. 
The main advantage of this design is to reach the requirements only through a
single spherical mirror, with the off-axis performance greatly improved by the
front correcting plate, with almost no chromatic aberrations and with UV
transmission enhanced with respect to a refractive optics. Having an $f/\# =
0.7$, the almost spherical photo-detector is small, implying overall mass
saving. Design studies have used a 5~m entrance pupil diameter, but they can be
somehow easily scaled to the desired dimension. The field of view is
$±25^{\circ}$ (half-angle), and the obscuration is limited. 

Protection against stray light is crucial: beside a light shield covering the
lateral side, a front baffle is being studied. The mirror is deployed with a
series of petals around a central structure. The design is being optimised to
use the biggest monolithic focal surface that will fit in the Ariane 5
fairing. Because of the size and of the difficulty to control temperature
gradients, the optical surfaces (both the primary mirror and the corrector
plate) must be actively controlled. The coupling between the thin optical
surface of the primary and the stiff lightweight support structure is made
through an array of actuators for the adjustment of the optical surface via
active control: this is necessary for improving the optical performance, for
in-orbit alignment but also for compensating thermoelastic  deformations of the
support. In the current baseline design 15 kg/m$^2$ 1-mm thick
Zerodur and Carbon Fiber Reinforced Plastic is used, for the supporting back plane.

\subsubsection{The Focal Surface.}
 The overall structure of the focal surface is being designed to follow a
 modular scheme. It consists of small autonomous functional units
 (elementary-cells) assembled in larger modules (photo-detector
 Modules). Modules are independent structures tied to each other by a common
 support structure and having a shape determined by the layout of the focal
 surface. The modular approach is crucial to allow sharing of resources, like
 supporting structures, power lines, cables, connectors and electronic, among
 sensors. The requirements for the photo-detector modules are: 1.) Capability to
 measure at single photo-electron level; 2.) Good charge resolution;  3.) High
 photon detection efficiency $\epsilon>0.6$; 4.) Good time resolution (two-pulse
 separation $2\div3$~ns); 5.) Pixel size $L_{pixel}\sim5$~mm; 6.) Low power
 consumption $<10$~mW~cm$^{-2}$; 7.) Low dark current (much less than
 background) and 8.) life-time of more than ten years. 

Multi-Anode Photo-multipliers (PMT), Flat Panel PMTs and Geiger-mode Avalanche
Photo Diodes, also called Silicon Photo-Multipliers (SiPM), are currently
investigated as \SEUSO possible sensors. Potential problems of PMTs are the
limited photon detection efficiency, non homogeneity of the response, relatively
high power consumption, poor flexibility in the design, and a rather weak and
relatively heavy mechanical structure. Avalanche Photo Diodes can be arranged in
large arrays which consists of about $10^3$ pixels. Large size photo diodes
arrays are the baseline sensor of \SEUSO. Their most attractive feature is the
potentially high photon detection efficiency ($>0.5$) and the capability of
single photon counting. Compactness of size and volume, low bias voltages, very
high gain of $10^6\div10^7$, insensitivity to magnetic fields, low power
consumption are other advantages. Larger size arrays, $5\times5$~mm${^2}$ or
better $10\times10$~mm$^{2}$, with high detection efficiency, low dark-counts,
low crosstalk and enhanced sensitivity in UV and blue light are required for
\SEUSO. Currently, arrays of size $5\times5$~mm${^2}$, micro-pixel size enlarged
to $100\times100 \mu$m$^2$, and photo-detection efficiency of 50\% at
$\sim500$~nm are being developed by the Semiconductor Laboratory of Max Planck
Institute for Physics~\cite{Teshima07}. Dark rate is $\sim0.5\div2.0$~MHz/mm$^2$
at room temperature and can be reduced by one or two orders of magnitude by
cooling the temperature down to $T\sim(-10)\div(-30)^{\circ}$C. Hamamatsu
Photonics has developed $1\times1$~mm$^2$ arrays which employs inverse polarity
for avalanche photo-diodes (p-on-n structure). This inverse structure enhances
the sensitivity to UV and blue light. Drawbacks are the narrow range of
operational bias voltages and high optical crosstalk between micro-pixels. INFN
has developed Multi-avalanche photo-diodes arrays of type n-on-p  from 1 to 16
mm$^2$. The current devices are optimised for blue light, with a 50\%
geometrical factor on 50x50 micro-pixel. Cross talk below 1\% and excellent
timing resolution (50 ps for single photon counting) have been obtained.
Back-Illumination-Drift avalanche photo-diode arrays are also being developed. A
photo-detection efficiency of 85\% or more in the range $330\div400$~nm could be
reached. Due to the large drift volume, dark current and cross talk may be
relatively higher compared to other systems.

%================================================================================

\subsubsection{Electronics and Trigger.}
The expected signal from an EAS is a track, a list of space-time correlated hits on the focal surface. Each hit is a bunch of photons in a pixel ($\sim 10$~photons/$\mu$s near the energy threshold; up to a few thousands photons/$\mu$s at ultra high energies). The expected background is several photons/$\mu$s, with significant variations. Typical EAS have a length from about  ten up to a hundred pixels. The electronics and the trigger system must be capable to sustain the high rate due to night-glow background and be tolerant to the enormous signals generated by lightings and human activities. As the background is variable in space and time along the orbit, we plan an on-board threshold setting system and a background subtraction system.  

Front end electronics will be a custom ASIC highly integrated with the sensors. it has to provide pre-amplification, shaping, and photon counting capability, with a programmable and self-adjusting threshold, background subtraction and zero suppression. To cope with luminous events (highest energy cosmic rays, Cherenkov reflection, luminous background sources) the front-end electronics must also provide charge integration. Eventually a track finding logic, in order to search for EAS like events, is necessary. At very low energy near threshold the expected signal is of the same order of magnitude of the background. With these small numbers, it is advisable to count the photons with a suitable preamplifier-discriminator-counter chain and identify the signal by putting a threshold on the counter, which should be periodically reset by a system clock. In this way, with a suitable on-chip logic, the system could have a self-adjusting threshold and subtract the expected background from each hit. We define the periodic reset clock a Gate Time Unit (GTU). Typical on-board programmable values of this gate range in $500$~ns $\div$ 2~s, depending on the operating conditions. The single photon counting technique will finally provide the number of collected photons for each GTU and for each pixel, allowing full reconstruction of energy and direction of the EAS. More details can be found in~\cite{Catalano2008} and references therein. The trigger system has to provide a fast trigger capable to manage several hundreds of thousands of channels. It has to be selective in order to tag the EAS signal while rejecting the background in an efficient way. The subtraction of the fluctuated background signal will be implemented in real time, making use of the Poisson property of systems based on counting. 

\subsubsection{The Monitoring System.}
The main purpose of the atmospheric monitoring system is required to characterize the earthÕs atmosphere inside the field of view of the instrument. The measurement objectives are 1.) mapping of the opaque clouds and determination of the cloud-top altitude; 2.) mapping of the sub-visible clouds and determination of their attenuation. The present concept for the monitoring system is based on the combination of several elements. An Infra Red camera will ensure mapping of the horizontal cloud coverage as well as the inputs for estimation of the cloud top with bias. The LIDAR measurements of the cloud-top altitude will be performed in several selected directions with high-precision and will be used to correct the bias in the IR camera. The proposed LIDAR will be a back-scatter type using wavelength in the UV spectral range, coinciding with the wavelength EAS fluorescence light. In this way it is possible to use the telescope as LIDAR receiver, where only several of the detectors will be used for back-scatter signal detection.  A calibration of the efficiency of the Instrument, using the molecular back-scatter signal and/or the signal scattered from cloud top and surface would be also possible as well as and evaluation of the albedo of the sea and land-mass surface. Details are found in~\cite{bi:SEUSO}.

%================================================================================
\subsection{Potentially critical issues affecting feasibility and performance}

Several critical points have been identified in this challenging project.
Only a full feasibility study, the so called phase-A study of space mission, can address them.

\begin{itemize}

\item
The deployable optics is a very challenging engineering task. However large
optics are highly desirable for many other future space applications. \SEUSO will benefit from other similar projects~\cite{bi:SO}. Moreover, a deployable catadioptric system has been studied by ESA in the context of an earth-looking LIDAR telescope. Deployability of the optics was also studied by NASA in the context of the OWL concept OWL~\cite{OWL}.

\item
The total ackground, including the random night-glow background, is very high: an online
subtraction, essential also to reduce the fake trigger count-rate, must be implemented.

\item
The observed field of view is continuously changing: a continuous monitoring and recording
of the relevant atmospheric parameters is required. The proposed concept is being validated through end to end specific simulations.  

\item
Orbit optimisation is dictated by several requirements: observational energy range, reduction of man-made
background, atmospheric phenomena, maximum night vs. day exposure, repetitive passages above specificed
ground-sources, very large drag coefficient and off-geometrical center Center of Mass.

\item
Optimal stray-light control of the large field of view as well as Photo-Detector protection from intense light (via attitude control and/or a shutter) are critical aspects of the current study.

\item
Data-handling, calibration and alignment for one million channels on orbit .

\end{itemize}

The demanding requirements have an impact on resources. 
A careful experiment optimisation is required which needs to
collect as much as possible information during the phase of mission conception
and design. A well defined road-map with intermediate steps is required~\cite{bi:RoadMap}.

%================================================================================
\section{Instrumental requirements and expected performance}
\label{pa:perfo}

The \SEUSO mission is an enlarged version of EUSO mission~\cite{bi:RedBook}. It improves the performance of EUSO by increasing the aperture and by exploiting novel technologies. A detailed analysis, leading to the results summarised in this paper can be found in~\cite{bi:SEUSO} and in~\cite{bi:ArXiv}. The current scientific scenario calls for an experiment capable to detect weakly interacting particles, with an order of magnitude lower energy threshold and an order of magnitude larger instantaneous geometrical aperture, with respect to EUSO, operating long enough to get an exposure greater than feasible on Earth. 

To increase the instantaneous geometrical aperture the height of the
orbit can be increased. This makes the signal fainter thus requiring an
even larger photon collection capability. The orbital parameters must
then be chosen to optimise the performance. In particular, to extend
the observational energy an elliptical orbit is considered.

The basic parameter which drives the performance is the photon
collection capability. It depends on the optics entrance pupil diameter
(optics aperture) and on the total photon collection efficiency. The
optical efficiency can be improved by using a catadioptric
optical system with a slightly reduced field of view, $\gamma_M = 20\degr \div
25\degr$, to get a better average optical efficiency. The gain with
respect to EUSO is $\approx 1.5$ assuming an optical
efficiency $\gtrsim 0.7$. The decreased instantaneous
geometrical aperture can be recovered with higher orbits. The photon
detection efficiency can improve by a factor $\approx 4$, thanks
to the newly developed Geiger-mode APD (GAPD) sensors which feature a
much larger quantum efficiency and a higher filling factor.

The optics aperture is the only sizable parameter which depends on mission
constraints. The maximum value of the entrance pupil diameter is chosen under the constrains of a non-deployable focal surface. The reflective deployable optical system has a $f/\# = 0.7$, with a goal of $f/\# = 0.6$. Using this assumption a factor $\approx10$ improvement with respect to EUSO is expected.

These figures alone give a factor $\approx 60$ improvement in the total
photon collection capability.  If one accounts for a perigee height twice larger
than EUSO, one still has more than a factor ten with respect to EUSO, allowing to bring the
total EAS detection efficiency down to $ E_{th} \simeq 10^{19} $ eV, as the EUSO
efficiency plateau was reached at about $ 10^{20}$ eV.  
We assume an elliptical orbit with apogee in the range $ r_A = ( 800\div1200 )$ km and perigee as low as possible, compatibly with constrains like atmospheric drag. The
perigee is currently assumed to be in the range $ r_A = ( 600 \div 1000 )\um{km} $. The
current baseline is: $ r_P = 800 \um{km} $ and $ r_A = 1100 \um{km} $.
The EAS triggering and reconstruction efficiency is another factor which will be improved by exploiting
the increased number of photons. 

The goal is to achieve an angular granularity of $ 0.04\degr $
corresponding to $\Delta \ell \approx 0.7 \um{km} $ when observing at half-FIELD from an
orbital height $(r_P+r_A)/2$.  The resulting Photo-Detector pixel size is $d=4 \um{mm}$ 
and the resulting number of channels is 1.2 million. The resulting orbit
averaged instantaneous geometrical aperture is $A_{eff} \approx 2\times10^6$~km$^2$~sr. 
This might be increased further by choosing a higher apogee, if advantageous. The outline of the estimations leading to the above instrumental
requirements from the scientific requirement are summarised in~\cite{bi:ArXiv}. 
The instantaneous geometrical aperture can also increase if the apparatus is
tilted with respect to the local nadir by some angle $\alpha_{tilt}$. To estimate the area observed at ground
with a tilted apparatus a Monte-Carlo integration has been performed in~\cite{bi:ArXiv}.  For $\alpha_{tilt}\geq30\deg$ the instantaneous geometrical aperture can be increased up to a factor $3\div5$. The drawbacks of the tilting mode however should be taken into careful consideration. Looking at the far extreme of the field of view,
drastically increase the role of atmospheric absorption implying that the effective energy threshold in the far part of the field of view increases. Also angular resolution at the far extreme becomes worse unless the pixel size is reduced. Excellent stray-light control is also required. Tilting, together with the large field of view, might affect the duty cycle, as the large area observed at Earth would more often include
day-time areas. The main \SEUSO baselines parameters and design goals, resulting from the
previous analysis are summarised in table~\ref{ta:params}. 

\begin{table}[htp]
\begin{center}
{\begin{tabular}{ll}
\hline
\multicolumn{2}{c}{\textbf{The main \SEUSO baselines parameters and design goals}}
\\
\hline
\hline
\multicolumn{2}{c}{\textbf{Main physical Parameters}}
\\
\hline
Operating Wavelength Range (WR)						
& 
	$\{330 \um{nm} \lesssim \lambda \lesssim 400 \um{nm}\}$
	\\
Background (in WR)~\footnote{at $\approx (r_P+r_A)/2$ height.}	
&
	$ \sci{(3 \div 15)}{11} \lumin $	
	\\
Average atmospheric transmission (in WR)				
&   
	$ K_{atm} \gtrsim 0.4$ 
	\\
\hline
\multicolumn{2}{c}{\textbf{Orbital Parameters}}
\\
\hline
Orbit perigee								
& 
	$r_P \simeq 800 $ km
	\\
Orbit apogee								
& 
	$r_A \simeq 1100 $ km
	\\	
Orbit inclination							
& 
	$ i \approx ( 50\degr \div 60\degr ) $ 
	\\								
Orbital period								
&
       	$ T_0 \simeq 100 $ min						
	\\									
Velocity of the ground track                                            
&	
	$v_{GT} \simeq 7.5 $ km/s
	\\									 
Pointing and pointing accuracy					
&
	Nadir to within $\Delta \xi \simeq 3\degr$			
	\\								
\hline
\multicolumn{2}{c}{\textbf{Satellite Parameters}}
\\
\hline
Satellite envelope shape
&
	Frustum of a cone
	\\
Diameters
&
	$D_{MAX} \simeq 11 $ m and $D_{MIN} \simeq 7 $ m.
	\\
Length
&
	$ L \simeq 10 $ m
	\\
Operational Lifetime
&
	$ (5 \div 10) $ years
	\\

\hline
\multicolumn{2}{c}{\textbf{Main instrument parameters and requirements}}
\\
\hline
Type
&
	Deployable catadioptric system
	\\
Main mirror
&
	$D_M \simeq 11$ m
	\\
Entrance pupil and corrector plate
&
	$D_{EP} \simeq 7$ m
	\\
Angular granularity
&
	$ \Delta \ell \approx 0.7 $ km at the Earth
	\\
Optics throughput
&
	$\epsilon_{O} \gtrsim 0.7 $
	\\
$ f/\# $
&
	$ \approx 0.6 $
	\\
Optics spot size diameter on the FS     
& 
	$ 3 \um{mm} \div 5 \um{mm} $   
	\\
Instrument Field of View, half-angle:
&
	$\gamma_M = 20\degr \div 25\degr$
	\\
Total length of the optics
&
$ \approx 9 $ m
\\
Focal Surface size (diameter)
&
	$ \approx 4 $ m
	\\
PDE
&
	$ \epsilon_{PDE} \gtrsim 0.25$
	\\
Number of detector channels
&
	$ \approx 1.2 $ million
	\\
Size of the pixels on the PD
&
	$ \approx 4 $ mm
	\\
Photo-Sensor
&
	GAPD
	\\
Power consumption
&
	less than $ 2 $ mW per channel
	\\
\hline
\multicolumn{2}{c}{\textbf{Main Performance Parameters and requirements}}
\\
\hline
Low Energy Threshold
&
	$ E_{th} \approx \scix{19} \um{eV}$.
	\\
Instantaneous geometrical aperture
&
	$ A_G \approx \sci{2.0}{6} \um{km^2 sr}$ 
	\\
Statistical error on the energy measurement
&
	$ \Delta E / E \approx 0.1 \; @ \; E \approx \scix{19} \um{eV}$
	\\
Angular resolution on the primary direction
&
	$ \Delta \chi \approx ( 1\degr \div 5\degr)$
	\\
Observation Duty cycle
&
	$ \eta \approx ( 0.1 \div 0.2 )$
	\\
\hline
\multicolumn{2}{c}{\textbf{Main budgets} (at the present level of knowledge)}
\\
\hline
Mass
&
	5 ton
	\\
Power
&
	5 kW
	\\
Telemetry
&
	20 Gbit/orbit
	\\
\end{tabular}}
\label{ta:params}
\end{center}
\end{table}

%================================================================================
\section{Conclusions}
In this paper we have described the science rationale, the scientific and instrument requirements, the conceptual design of a next-generation space mission for the exploration of the ultra high energy Universe. Although the mission appears technologically feasible we are aware that  several critical issues which range from background assessment, to technology readiness of the components, to the management of a complex one million channels readout, and to the optimisation of the atmosphere monitoring system should be addressed. 
Such a challenging space-based experiment certainly requires a number of developments to optimise the design, qualify the observational technique, perform preliminary measurements and test critical
parts. In particular measurements of air fluorescence yield, of the Cherenkov albedo, and of the background observed from space are crucial. Technological tests via stratospheric airplane flights and/or
balloon flights can help in optimising the mission parameters. However a small
instrument on-board a mini-satellite could provide a detailed characterisation
of the background and of the duty cycle in space conditions and a test of
critical technological items. As well a measurement of the light level far
off-nadir for stray-light control could be obtained. So the road ahead is not
easy. However we believe that such a mission tough challenging is essential to
unveil at the end of the decade the still unexplored ultra high energy Universe.

%================================================================================
%================================================================================
\ack
This paper is largely based on the work done by the EUSO and the \SEUSO collaboration which is reported in details in~\cite{bi:SEUSO} and~\cite{bi:RedBook}. So we wish to warmly thank all members of these collaborations. A special thought goes to Livio Scarsi. Without his restless and enthusiastic efforts the space
based research for ultra high energy particles would not have reached its present exciting state.%================================================================================


\section*{References}

\begin{thebibliography}{99}

\bibitem{Auger1} 
Abraham J \textit{et al }(The Pierre Auger Collaboration) 2008 \textit{Phys. Rev. Lett.} \textbf{101} 0611101.

\bibitem{bi:SEUSO} 
Santangelo A, Petrolini P \textit{et al} 2007
\textit{The \SEUSO proposal} \url{http://astro.uni-tuebingen.de/groups/uhe/cv-proposal/S-EUSO-proposal-29.06.2007.pdf}.

\bibitem{Watson08} 
Watson A A 2008 \textit{Nucl. Instr. Meth.} \textbf{A588} 221.

\bibitem{bi:JEMEUSO} 
Takahashi Y \textit{et al} 2007 \textit{J. Phys. C.} \textbf{65} 012022 \url{http://jemeuso.riken.jp/}.

\bibitem{Blasireview} Blasi P 2005 \textit{Mod. Phys. Lett.} \textbf{A20} 3055.

\bibitem{GZK} 
Greisen K 1966 \textit{Phys Rev. Lett.} \textbf{16} 748;
Zatsepin G T, Kuz'min V A Zh 1966 \textit{Eksp. Teor. Fiz.} \textbf{4} 114;
Zatsepin G T, Kuz'min V A Zh 1966 \textit{JETP Lett.} \textbf{4} 78.

\bibitem{Stecker68} 
Stecker F 1968 \textit{Phys. Rev. Lett.} \textbf{21} 1016.

\bibitem{Linsley63} 
Linsley J 1963 \textit{Phys. Rev. Lett.} \textbf{10} 146.

\bibitem{NaganoWatson2000} 
Nagano M and Watson A A 2000 \textit{Rev. Mod. Phys.} \textbf{72-73} 689.

\bibitem{Bird93} 
Bird D J \textit{et al} 1993 \textit{Phys. Rev. Lett.} \textbf{71} 3401.

\bibitem{Takeda03} 
Takeda M \textit{et al} 2003 \textit{Astropart. Phys.} \textbf{19} 447.

\bibitem{Abu04} 
Abu-Zayyad T. \textit{et al} (HiRes Collaboration) 2004 \textit{Phys.Rev.Lett.} \textbf{92} 151101.

\bibitem{HiRes08} 
HiRes Collaboration 2008 \textit{Phys. Rev. Lett.} \textbf{100} 101101.

\bibitem{Takeda99} 
Takeda M \textit{et al} 1999 \textit{Astrophys. J.} \textbf{522} 225.

\bibitem{Abbasi04} 
Abbasi \textit{et al} 2004 \textit{Astrophys. J.} \textbf{610} L79.

\bibitem{Farrar05} 
Farrar G R \textit{et al} 2006 \textit{Astrophys. J.} \textbf{642} L89.

\bibitem{Auger2} 
Abraham J \textit{et al} (The Pierre Auger Collaboration) 2007 \textit{Science} \textbf{318(5852)} 938.

\bibitem{Kashi08} 
Kashti T and Waxman E 2008 \textit{JCAP} \textbf{05} 006. 

\bibitem{Ghisellini08} 
Ghisellini G \textit{et al} 2008 \textit{Preprint} astro-ph/0806.2393.

\bibitem{Berezinsky05} 
Berezinsky V 2005 \textit{Preprint} astro-ph/0509675.

\bibitem{Kusenko2000} 
Kusenko A. and Weiler T. 2002 \textit{J. Phys. Rev. Lett.} \textbf{88} 161101.

\bibitem{Fargion1999} 
Fargion D, Mele B and Salis A 1999 \textit{Astrop. J} \textbf{157} 725.

\bibitem{Palomares2006} 
Palomares-Ruiz S, Irimia A and Weiler T J  2006 \textit{Phys. Rev. D} \textbf{73} 083003.

\bibitem{Augerphoton} 
Abraham J \textit{et al} (The Auger Collaboration) 2008 \textit{Phys. Rev. Lett.} \textbf{100} 211101.

\bibitem{Linsley82} 
Linsley J 1982, \textit{Proc. of the Workshop on Very High Energy Cosmic-Ray Interactions} University of Pennsylvania, eds. Cherry M L, Lande K and Steinberg R I, 476.

\bibitem{Takahashi09} 
see the contribution from Taskahashi Y (JEM-EUSO Collaboration) in this Focus Issue.

\bibitem{Santangelo2005} 
Santangelo A, Petrolini A and Plagnol E 2005 Physics and astrophysics at
Ultra High Energies - A Cosmic Vision theme for the search of UHE CR and
neutrinos from space \textit{Proc. of 39th ESLAB Symposium}, Favata \textit{et al} Eds, Noordwijk, The Netherlands.

\bibitem{Kakimoto} 
Kakimoto F \textit{et al} 1996 \textit{Nucl. Instr. Meth.} \textbf{A372} 527.

\bibitem{Nagano} 
Nagano F \textit{et al} 2004 \textit{Astropart. Phys.} \textbf{22} 235.

\bibitem{bi:ArXiv}
Pallavicini M, Pesce R, Petrolini A and Thea 2008 The observation of Extensive Air Showers from Space
\textit{Preprint} astro-ph/0810.5711v1.

\bibitem{Barbier} 
Barbier L M \textit{et al }2005 \textit{Astropart. Phys.}\textbf{22} 439.

\bibitem{Garipov} 
Garipov G K \textit{et al} 2005 \textit{JEPT Letters} \textbf{82(4)} 185.

\bibitem{Berat2003} 
Berat C \textit{et al},  2003 The light of the night sky in EUSO: Duty cycle and Background \textit{Proc. 28th ICRC} Tsukuba, Japan.

\bibitem{Optic1} 
Mazzinghi P et al. 2006 An ultra-lightweight, large aperture, deployable telescope for advanced LIDAR applications \textit{Proc. 6th Int. Conf. on Space Optics}, Noordwijk, The Netherlands (ESA SP-621).

\bibitem{Optic2} 
Simonetti F, Romoli A, Mazzinghi P and Bratina V 2006 \textit{Optical Engineering} \textbf{45} (05).

\bibitem{OWL} 
Stecker F W \textit{et al} 2004 \textit{Nucl. Phys. Proc. Suppl.} \textbf{136C} 433.

\bibitem{Teshima07} 
Teshima M \textit{et al} 2007 \textit{Preprint} astro-ph/0709.1808.
 
\bibitem{Catalano2008}  
Catalano O, Maccarone M C and Sacco B 2008 \textit{Astr. Part. Phys.} \textbf{29} 104.

\bibitem{bi:RedBook} Scarsi L, Catalano O, Santangelo A, Petrolini A \textit{et al} 2003
\textit{The EUSO Redobook} Internal note EUSO-PI-REP-002.


\bibitem{bi:SO}
C. F. Lillie and W. B. Whiddon,
\textit{Deployable Optics for Future Space Observatories}, presented at
Space 2004 Conference and Exhibit, 28-30 September 2004, San Diego,
California, AIAA 2004-5894;
\url{http://pdf.aiaa.org/preview/CDReadyMSPACE2004_1014/PV2004_5894.pdf}.\\
C. F. Lillie, \textit{Large deployable telescopes for future space observatories},
UV/Optical/IR Space Telescopes: Innovative Technologies and Concepts II. Edited by MacEwen, Howard A. Proceedings of the SPIE, Volume 5899, pp. 108-119 (2005).


\bibitem{bi:RoadMap}
Petrolini A 2008 \textit{Nucl. Instr. and Meth.} \textbf{A588} 201.


\end{thebibliography}
\end{document}